# Proposal of thermal neutron flux monitors based on vibrating wire

S.G.Arutunian,[1] J.Bergoz,[2] M.Chung,[3] G.S.Harutyunyan,[1] E.G.Lazareva[1]

[1] *Yerevan Physics Institute, Alikhanian Br. St. 2, Yerevan, 0036, Armenia*

[2] *Bergoz Instrumentation, 156, rue du Mont Rond, 01630, France*

[3] *Ulsan National Institute of Science and Technology, gil 50, Ulsan, 689-798, Korea*

**Abstract**

Two types of neutron monitors with fine spatial resolution are proposed based on vibrating wire. In the first type, neutrons interact with the vibrating wire, heat it, and lead to the change of natural frequency, which can be precisely measured. To increase the heat deposition during the neutron scattering, use of gadolinium layer which has the highest thermal neutron capture cross section among all elements is proposed. The second type of the monitor uses vibrating wire as a "resonant target". Besides the measurement of beam profile according to the average signal, the differential signal synchronized with the wire oscillations defines the gradient of beam profile. Spatial resolution of the monitor is defined by the diameter of the wire.

## 1. Introduction and Overview

### 1.1. Introduction

Many research centers in the world use neutrons as probes to investigate diverse properties of a wide range of materials. Neutrons are excellent for probing materials on the molecular level – everything from motors and medicine, to plastics and proteins [1]. Since neutrons are scattered by atomic nuclei, they provide information on the structure and dynamics of atoms and molecules over a wide range of length and time scales. Because of the presence of magnetic moment, neutrons can also be used for the study of the magnetic structures and dynamics of materials.

Neutrons have found interesting uses in medicine, both in the treatment of cancer and in the development of ultra-sensitive analytical techniques for the study of man's internal and external chemical environment [2]. This approach combines thermal neutrons and chemical markers. There are a few centers specialized in neutron therapy [3], where neutrons are particularly generated from low energy cyclotrons. Control of the spatial distribution of the beam and its intensity in this field of usage is vital.

The neutron is a powerful probe for the study of condensed matters (disordered, amorphous, glass, crystalline, non-equilibrium, and nanocomposite materials) in the world around us. Neutron scattering gives detailed information about atomic level structure and dynamics, that is, where atoms are located and how they are moving. Neutrons used in the scattering experiments have wavelengths similar to the atomic spacing, allowing the structures of materials to be studied by diffraction on the length scales from atomic dimensions to macromolecular scales. At the same time, neutrons have energies similar to those of atomic processes, such as molecular transitions, rotations, vibrations, and lattice modes. Good introduction to the theory, techniques, and applications of the neutron scattering is given in the Teaching materials of the 13th Oxford School on Neutron Scattering [4]. So there is a need for devices capable of real-time aquisition/tracking of the beam intensity.

Most of the existing neutron sources are based on nuclear reactors. Advantages: high time averaged flux, mature technology (source and instruments). Disadvantages: licensing (cost/politics), no time structure [5]. The other types of the neutron sources are so-called spallation facilities, where neutrons are produced by bombardment of heavy nucleus target by high-power proton beam. Spallation sources can be either pulsed or continuous. Several new neutron sources are planned to be operational in the future.

More detailed description of the neutron sources worldwide is given in Sec. 1.2.

Of great importance is to form neutrons in a beam with high flux, and well-identified and precisely-controlled parameters of intensity, divergence, geometrical coordinates and sizes. To reach these requirements, a new generation of neutron sources has been developed based on particle accelerators and spallation technology. So far, neutron beam's flux intensity reaches to $10^{13}$-$10^{15}$ n/cm$^2$/s. The new task is to increase the intensity of the neutron beams further. For example, the European Spallation Source (ESS) is designed to provide around 30 times brighter neutron beams than the existing facilities today [6, 7].

Instrumentation for the neutron beams needs a variety of detectors and monitors for measurements and controls. All known detectors for slow neutrons today are based on the conversion of neutrons into charged particles (see e.g. [8]).

After this conversion, the following technologies are normally used: gas proportional counters, ionization chambers, scintillation detectors, and semiconductor detectors. As conversion materials, He, Li, B, and Gd are often used due to their high values of nuclear reaction cross-sections. In most cases information on the spatial distribution of the neutron beam is also necessary. However, the above mentioned approaches provide this information rather implicitly and roughly.

In this proposal, as a source of detectable signal, we suggest to use immediate interaction of neutrons in media that leads to heat deposition in a wire – a core piece of detector. Indeed, we intend to combine two unique properties – the unprecedented sensitivity of the natural frequency of a clamped vibrating wire to the wire temperature, and remarkable ability of some gadolinium isotopes for neutron capture. We propose to measure temperature increase of the wire containing gadolinium isotopes. This temperature increase occurs when neutrons penetrate the wire and deposit some energy into the wire. Due to wire's small diameter, this method can resolve local neutron beam density along the wire. Vibrating wire neutron monitor (VWNM) with composite wires and wide dynamic range will be developed, aiming for a precise spatial resolution profiling of the high flux neutron beams from specialized neutron sources (research reactors and spallation source) on the inlet of neutron based instruments. Different from the existing neutron measurement principles, the proposed method has high spatial resolution, depending on the wire diameter.

The second method proposed here is so-called "Resonant target" vibrating wire monitor [9]. Measurements of gamma-rays which arise from neutron scattering on the oscillating wire atoms (e.g. gadolinium) in synchronism with the wire's oscillation frequency can provide a prompt response of the scattering process. The differential signal in wire's maximum deviations during oscillation process can provide fast information on the beam profile gradient. Secondary particles and radiation arise from the neutron scattering depend on vibrating wire's oscillation phase, and thus can be used for recovering beam flux density by measurements synchronized with vibrating wire's oscillation frequency. Transverse beam profile measurement by this method was tested with lightening the oscillating wire by a laser [9]. Use of sensitive gamma-ray detectors can improve the resolution of the neutron beam profile measurements compared with the direct thermal measurements by vibrating wire monitor only. So Resonant target Vibrating wire neutron monitor (RT-VWNM) with composite wires and subsidiary gamma-ray detectors will be developed aiming for the fast profiling of the low flux neutron beams and their transverse gradients.

Both methods are described in detail in the corresponding sections below

The vibrating wire technology was first applied for the measurement of low current electron beams in the injector of Yerevan synchrotron [10]. The principle of the operation was to measure the frequency shift of natural frequency of stretched wire exposed to the beam. Even a small number of electrons penetrating the wire were enough for producing a detectable change of frequency. Since then, many other applications of the vibrating wires have been proposed. For example, it was demonstrated that very high sensitivity as low as mK could be achieved in the measurement of temperature shift. This feature was used for beam halo measurements (see e.g. [11-13]).

So being developed as a result of project fulfillment, monitors VWNM and RT-VWNM can be widely used for all applications of neutron beams. For example, the VWNM can be used as a precise monitor with excellent spatial resolution for high flux neutron beams of specialized neutron sources with multibranch infrastructure of numerous instruments for material research. And the RT-VWNM can be used as a robust and reliable instrument with also excellent spatial resolution for low flux neutron beam diagnostics (e.g. for centers of neutron therapy). Specialized multiwire VWNM with capability of rotating along the beam axis can be used for the recovery of complicated 2D profiles of large cross-section neutron beams in neutron tomography, imaging and radiography.

VWNM's can be used in 18 MeV Cyclotron (Cyclone 18) of Yerevan's oncological center for direct beam profile measurements in medical treatment. Another area of usage can be diagnostics of neutron beam planned to be generated at Cyclone-18 for use in a broad class of studies and experiments (engineering of materials, biological, chemical, and physical systems investigations, astrophysics, nuclear physics, and material science) [http://armenianweekly.com/2014/12/30/new-cyclotron-yerevan-physics-institute/].

Preliminary experiments and tests is planned to perform on the $^{252}$Cf spontaneous fission neutron source accessible in Yerevan Physics Institute.

## 1.2. Review of neutron sources

### 1.2.1. World neutron sources

In Table 1 we present most of the existing neutron sources in the world with brief description of the source type and parameters, main characteristics of the neutron beam, and list of instruments available at the sources [14-35]. Some of the neutron sources will be described in more detail at the end of this section.

**Table 1. World neutron source**

| Source name | Source type and neutron beam parameters | Instruments |
|---|---|---|
| Oak Ridge Neutron Facility, High Flux Isotope Reactor HFIR | Reactor/ power 85 MW; Thermal neutron flux 2.6 x $10^{15}$ n/cm$^2$/s | Neutron imaging facility; Small-angle neutron scattering Diffractometer (flux on sample 2 x $10^7$ n/cm$^2$/s); Biological small-angle neutron scattering instrument; Cold neutron triple-axis spectrometer; Image-plate single crystal diffractometer (flux ~$10^7$ n/s/cm$^2$); Polarized triple-axis spectrometer; Fixed-incident-energy triple-axis spectrometer (flux at sample-2 x $10^7$ n/cm$^2$/s); Neutron powder diffractometer; Neutron residual stress mapping facility( 3 x $10^7$ n/cm$^2$/s); Wide-angle neutron diffractometer; Four-circle diffractometer(2.2 x $10^7$ n/cm$^2$/s) |
| Oak Ridge Neutron Facility, Spallation Neutron Source SNS | Spallation/ 1 GeV proton accelerator (1.4 mA); Different fluxes at different instruments in range of $10^7$ n/cm$^2$/s - $10^9$ n/cm$^2$/s | Nanoscale-ordered materials diffractometer, flux on Sample~(1 x $10^8$ n/cm$^2$/s); Backscattering spectrometer Spallation neutrons and pressure diffractometer; Magnetism reflectometer; Liquids reflectometer; Cold neutron chopper spectrometer; Extended Q-range small-angle neutron scattering Diffractometer(~$10^7$-$10^9$ n/cm$^2$/s); Engineering materials diffractometer (2.2 x $10^7$ n/cm$^2$/s In high-resolutionmode6.7 x $10^7$ n/cm$^2$/s in high-Intensity mode); Powder diffractometer; Macromolecular neutron diffractometer; Single-crystal diffractometer; Hybrid spectrometer, polarized; Neutron spin echo spectrometer; Al spectrometer; Fine-resolution fermi chopper spectrometer; Wide angular-range chopper spectrometer |
| Los Alamos Neutron Science Center LANSCE | Spallation/ 20Hz 800 MeV proton beam current 100 - 125 µA; 0.1- 600 MeV neutrons produced an unmoderated tungsten (Target 4), sub-thermal - 500 keV at the Lujan Center, tungsten target with proton 100 µA (Target 1) | Neutron powder diffractometer; High-intensity powder diffractometer; Neutron time-of-flight powder diffractometer; Single-crystal diffractometer; Low-Q diffractometer; Surface profile analysis reflectometer; Filter difference spectrometer; High-resolution chopper spectrometer; Detector for advanced neutron capture experiments |
| University of Missouri Research Reactor Center, Reactor MURR | Reactor/ power 10MW; Flux trap, peak 6 x $10^{14}$ n/cm$^2$/s | Triple –axis spectrometer; Neutron reflectometer |
| Canadian Neutron Beam Centre, Chalk River, Canada, Reactor NRU | Reactor/ power 125 MW; Peak thermal flux 3 x $10^{14}$ n/cm$^2$/s | High resolution powder diffractometer; Polarized beam triple-axis spectrometer; Reflectometer; Triple-axis spectrometer; Stress-scanning diffractometer; Magnet cryostat; Triple-axis spectrometer; Neutron spectrometer ancillary equipment; Source and main beam specifications; Image-plate diffractometer |
| Budapest Neutron Centre, Hungary | Reactor/ thermal power 18 MW, mean power density(39.7 kW/l); Flux 2.1 x $10^{14}$ n/cm$^2$/s | Three-axis spectrometer on neutron guide; Port used biological irradiation; Powder diffractometer; Material test diffractometer; Time-of-flight diffractometer; Small angle neutron scattering; Neutron induced prompt gamma spectrometer; Prompt gamma activation analysis; Polarized beam neutron reflectometer; Dynamic n/γ radiography, static radiography |

| Facility | Source | Instruments |
|---|---|---|
| St. Petersburg Neutron Physics Institute, Gatchina, Russia, Reactor WWR-M | Reactor/ power 18 MW; Flux $4 \times 10^{14}$ n/cm$^2$/s | Crystal-diffraction monochromator of neutrons; 48 countering powder diffractometer; Reflectometer; Small-angle diffractometers; Single-crystal diffractometer of polarized neutrons; Powder diffractometer; Spin echo spectrometer; Spin echo small-angle neutron scattering facility; Polarized neutrons reflectometer |
| Frank Laboratory of Neutron Physics, Dubna, Russia, Reactor IBR-2 | Reactor/ power 1850 MW; Flux $10^{16}$ n/cm$^2$/s | Neutron diffractometers; Small-angle scattering; Reflectometer; Inelastic scattering |
| Hi-Flux Advanced Neutron Application Reactor, Korea, "HANARO" or "KMRR" High-flux Reactor | Reactor/ power 30 MW; Peak thermal neutron flux $5.4 \times 10^{14}$ n/cm$^2$/s typical flux at port nose $2 \times 10^{14}$ n/cm$^2$/s | High resolution powder diffractometer; Small Angle Neutron Scattering Instrument; Vertical Neutron Reflectometer; Four-circle diffractometer; Prompt Gamma Neutron Activation Analysis; Neutron Radiography |
| Institute Laue-Langevin, Grenoble, France, High-Flux Reactor | Reactor/ thermal power 58.3 MW; continuous neutron flux in the moderator region: $1.5 \times 10^{15}$ n/cm$^2$/s | Powder diffractometers; Large-scale structure diffractometers; Time-of-flight spectrometers; Three-axis spectrometers; Single-crystal diffractometers; Reflectometers; High-resolution spectrometers |
| Helmholtz-Zentrum Geesthacht, Germany, Reactor GEMS at | Reactor/ light water moderated swimming pool reactor, 10 MW thermal power; $2 \times 10^{14}$ n/cm$^2$/s in the core | 3-axis spectrometer; Diffractometers; High magnetic field facility for neutron scattering; Reflectometer; Small angle neutron scattering instrument stress and Strain/textures; Time-of-flight spectrometer; Tomography/radiography |
| Paul Scherrer Institut PSI, Switzerland, SINQ | Spallation/ 590 MeV proton accelerator (current up to 2.3 mA); Continuous flux of about $10^{14}$ n/cm$^2$/s | Diffractometers; Small-angle scattering instruments; Reflectometers; Spectrometers; Non-diffractive instruments |
| IFE, Kjeller, Norway, Reactor JEEP-II | Reactor/ power 2MW; Neutron flux of $1.3 \times 10^{13}$ n/cm$^2$/s | High resolution gamma ray spectrometry |
| J. Stefan Institute, Slovenia Ljubljana, TRIGA MARK II Research Reactor | Reactor/ power 0.25 MW; Central thimble $10^{13}$ n/cm$^2$/s | Elastic and inelastic neutron scattering; Neutron dosimetry; Neutron radiography |
| Nuclear Physics Institute ASCR, Prague, Czech Republic, Reactor LVR-15 | Reactor/ power 10 MW; Maximum thermal neutron flux in the core $1 \times 10^{14}$ n/cm$^2$/s, maximum thermal flux in reflector $5 \times 10^{13}$ n/cm$^2$·s | Neutron powder diffractometer; Neutron optics diffractometer; High-resolution SANS diffractometer Multipurpose double axis diffractometer |
| Kyoto University Research Reactor Institute, Kyoto, Japan, Reactor KURRI | Reactor/ thermal power 5MW; Neutron flux in range of $8.2 \times 10^{13}$ - $3 \times 10^{12}$ n/cm$^2$/s | Special environment neutron diffractometer; Quasi-elastic neutron scattering; $^{10}$B-neutron capture therapy (BNCT) |
| Malaysian Nuclear Agency, Malaysia, Reactor Triga Puspati RTP, | Reactor/ Maximum thermal power 1MW, Pulsing Peak Power-1200MW (pulse width 11ms); Maximum thermal neutron flux $1 \times 10^{13}$ n/cm$^2$/s | Neutron radiography facility; Small angle neutron scattering facility |

### 1.2.2. Other types of neutron sources

Neutrons can be produced by spontaneous fission [36]. For example, $^{252}$Cf undergoes α-decay with a half-life of 2.65 years and n-decay with a half-life of 85.5 years. The average neutron energy is 2.14 MeV, and the decay rate is 2.34 x $10^{12}$ n/s/g. Such sources are produced by irradiation in high-flux reactors, in which neutron fluxes are in the range of ($10^5$ n/s to $10^9$ n/s) or ($10^8$ n/s to $10^{10}$ n/s) [37]. Another way of using radioactive materials is the neutron production through a secondary process. For example, in an α-n source such as $^{241}$Am/Be, $^{241}$Am undergoes α-decay; the α-particles are absorbed by beryllium, which then decays by neutron emission. This can be written as

$$^{241}\text{Am} \rightarrow {}^{237}\text{Np} + {}^{4}\text{He (5.6MeV)}, \qquad (1)$$

followed by

$$^{9}\text{Be} + {}^{4}\text{He} \rightarrow {}^{12}\text{C} + n \text{ (few MeV)}. \qquad (2)$$

The half-life is 433 years, and these sources can produce neutrons with the rate of $10^6$–$10^8$ n/s per 1 g of Am.

Higher intensities can be achieved by small accelerator-based neutron sources. Over the years, these sources have evolved so remarkably that compact portable sources are now commercially available from a range of vendors. They are normally based on the ''D-T'' reaction:

$$^{2}\text{H}(\sim 150 \text{ keV}) + {}^{3}\text{H} \rightarrow {}^{4}\text{He} + n(14:2\text{MeV}) \qquad (3)$$

Sealed tube sources with a typical length of 1 m and diameter of 10 cm, operating at a power of 0.5 kW, can produce up to 3 x $10^{10}$ n/s. At a distance of 1 m, this gives a flux of fast neutrons on the order of 2 x $10^5$ n/cm$^2$/s.

Besides, coaxial RF-driven plasma ion source for a compact cylindrical neutron generator has been developed. The single target coaxial neutron generator with dimensions of 26 cm in diameter and 28 cm in length is expected to generate a 2.4 MeV D-D neutron flux of 1.2 x $10^{12}$ or a 14 MeV D-T neutron flux of 3.5 x $10^{14}$ n/s [38].

Recently, the idea of coupling a sub-critical fission reactor and a DT-plasma device generating 14 MeV neutrons for the incineration and transmutation of long-lived isotopes has attracted increasing interest [39]. Such DT-plasma surrounded by fission blanket provides some advantages as compared to accelerator-driven systems, like spallation source.

A novel compact laser-driven neutron source with an unprecedented short pulse duration (<50 ps) and high peak flux (>$10^{18}$ n/cm$^2$/s), an order of magnitude higher than any existing source was reported [40]. In the experiments, high-energy electron jets are generated from thin (< 3 μm) plastic targets irradiated by a petawatt laser. These intense electron beams are employed to generate neutrons from a metal converter. Authors mentioned that the method opens venues for enhancing neutron radiography contrast and for creating astrophysical conditions of heavy element synthesis in the laboratory.

### 1.2.3. Neutron detectors

Since neutrons are neutral particles and do not ionize matters directly, it is impossible to detect them straight forwardly. It seems that techniques for detection of the magnetic moment of neutrons are too insensitive to use for the neutron detection. So most detection approaches rely on detecting the various reaction products. Any neutron-nucleus interaction leaves the nucleus in an excited level from which it decays by emitting gamma-rays. Prompt gammas are emitted within $10^{-13}$ s and are always associated with the neutrons moving in the matter. Prompt gamma-ray's energy depends upon the neutron energy. Typical spectral lines are measurable for each isotope (e.g. 2.2 MeV from H) and can be as high as 10-15 MeV. Nuclear reactions A(n,x)B leave the reaction products in excited levels from which they decay emitting particles α, β, etc. with variable half-lives [41].

*Gas Detectors*

Typical reaction involves a chamber filled with Helium 3:

$$n + {}^{3}\text{He} \rightarrow {}^{3}\text{H} + {}^{1}\text{H} + 0.76 \text{Mev} \qquad ()$$

with cross-section $\sigma = 5333(\lambda/1.8)$ barns ($\lambda$ is neutron De Broglie's wavelength in angstrom, $\lambda = 1.8$ Å for neutron energy of 26 meV). As a result about 25000 ions and electrons are produced per neutron [8]. In ionization mode of the gas detector, electrons drift to anode, producing a charge pulse and corresponding signal. In proportional mode, high voltage is applied so electron collisions ionize gas atoms producing even more electrons. In this case, gains up to a few thousands are possible.

*Scintillation Detectors*

Converting medium here is a scintillator that produces numerous photonsas a result of neutron reaction. Ina Li-based scintillator, following reaction occurs

n+$^6$Li->$^4$He+$^3$H+4.79MeV           (4)

with cross-section $\sigma = 940(\lambda/1.8)$ barns. As a result, about 7000 photons/n are produced in Li glass (Ce), about 51000 photons/n in (LiI (Eu)), and about 160000 photons/n in (ZnS(Ag)-LiF),respectively[8]. Photomultipliers are commonly used following the photon production.

*Semiconductor Detectors*

Most conventional semiconductor materials have very low probabilities of interacting with free neutrons. For example, the microscopic thermal neutron cross-section for naturally occurring silicon is ~ 2.24 barns, which means a thermal neutron would have to travel an average distance of ~ 8.6 cm (mean free path) before a scattering or capture event would occur [42].The way to overcome the low probability of neutron interaction with conventional semiconductor materials and improve detection efficiency is to integrate a layer of neutron reactive material into the semiconductor device architecture. These designs, commonly referred to as conversion layer devices, typically use Si or GaAs p-n junction, or Schottky diodes to separate electron-hole pairs generated from interactions with primary nuclear reaction products [43]. $^{10}$B and $^6$LiF are frequently used as conversion layer materials because of their stability, large thermal neutron cross-sections, and primary reaction products.

### 1.2.4. Neutron beam instruments and applications

There are many instruments that use neutron beams for the investigation of matter. Below we describe some of them [44].

**Neutron Diffractometers:** Study of lattice dynamics of crystalline, amorphous materials and liquids, Determination of structural parameters of crystalline materials with high precision, Determination of structural parameters of crystalline materials and nanosystems (lipid membranes, etc), real-time studies of chemical and physical processes, Determination of parameters of crystal and magnetic structure of materials as function of external pressures, In situ studies of macro- and microstresses in rocks, Studies of texture of geological samples (rocks, minerals), Determination of residual stresses in bulk industrial components and new advanced materials, Determination of parameters of crystal and magnetic structure of materials as function of external pressures

**Small-angle scattering:** Determination of structural characteristics (size and shape of particles, agglomerates, pores, fractals) of nanostructured materials and nanosystems, including polymers, lipid membranes, proteins, solvents, etc

**Reflectometry:** Determination of magnetization profile of layered magnetic nanostructures, studies of proximity effects in nanosystems, Determination of structural characteristics of thin films and layered nanostructures, Studies of surface and interface phenomena in soft and liquid nanosystems (magnetic fluids, polymers, lipid membranes)

**Inelastic scattering:** Study of lattice dynamics and structural parameters of molecular crystals, crystals with molecular ions, especially exhibiting polymorphism,

**Instrumental Neutron Activation Analysis:** The non-destructive version of NAA permit analysis of archaeological specimens (pottery, coins, etc.) for characterization, providing useful information about sources and trade routes, without destroying the specimen

**Radiography:** Neutron radiography (N-ray or NR) and X-radiography (X-ray) are complementary non-destructive testing techniques. In both cases, a form of radiation passes through the object being imaged, and then exposes a photographic film. Neutron radiographs and X-radiographs show different characteristics of the object imaged due to differences in neutron and X-ray interaction with the material that the object is made up of (see Figure 1 [45]). The images provide good examples of the differences between neutron radiography and X-radiography. Liquids, such as the lighter fluid shown in the lighter image, plastics, rubbers, ceramics and lubricants show up very well in neutron radiographs, while metal components show up well in X-ray images. Due to the relative invisibility of metals to neutrons, neutron radiography can be used to effectively image items encased in metal. The variation in images produced using neutron radiography and X-radiography makes the two complementary technologies, both very useful for particular applications. The internal structure, air pathways and blockages or inclusions in metal alloy turbine blades can be imaged very clearly using neutron radiography.

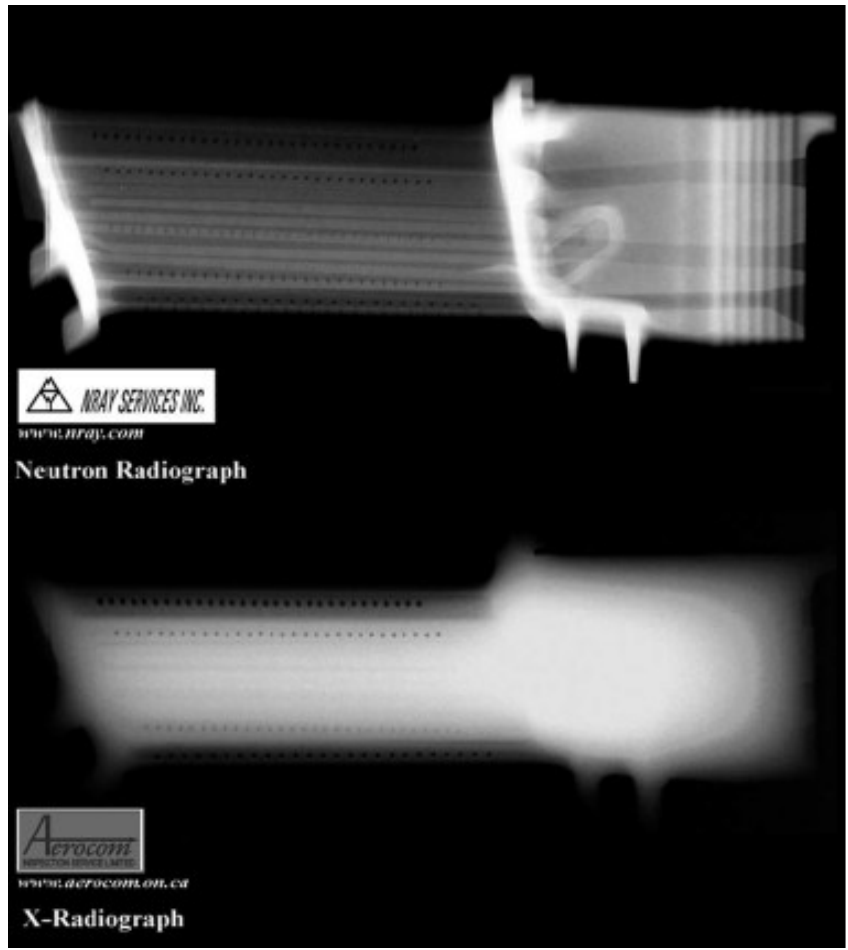

Figure 1. Turbine blade in neutron radiography (top view) and X-ray radiography (bottom view).

**1.2.5. Detailed characteristics of ILL HFR and HANARO, Korea**

The aim of our proposal is development of new instruments for neutron beam diagnostics with spatial resolution. The values of neutron beam fluxes in Table 1 actually present intensities in a few fixed points: in the core of reactors, in reflector (for reactor type sources) etc. Every neutron source has a lot of beamlines that extract and form the neutron beams with different range of energies, and guide them to the numerous stations with different instruments, so it is an urgent task to accumulate particular information at these transfers.

To imagine more clearly the complicated and manifold structure of specialized neutron sources, we present detailed information for two reactor-based neutron sources: ILL and HANARO.

In Figure 2 the main view of the beamlines, neutron beam types, and instruments in Institute Laue-Langevin (ILL) High-Flux Reactor are presented [46]

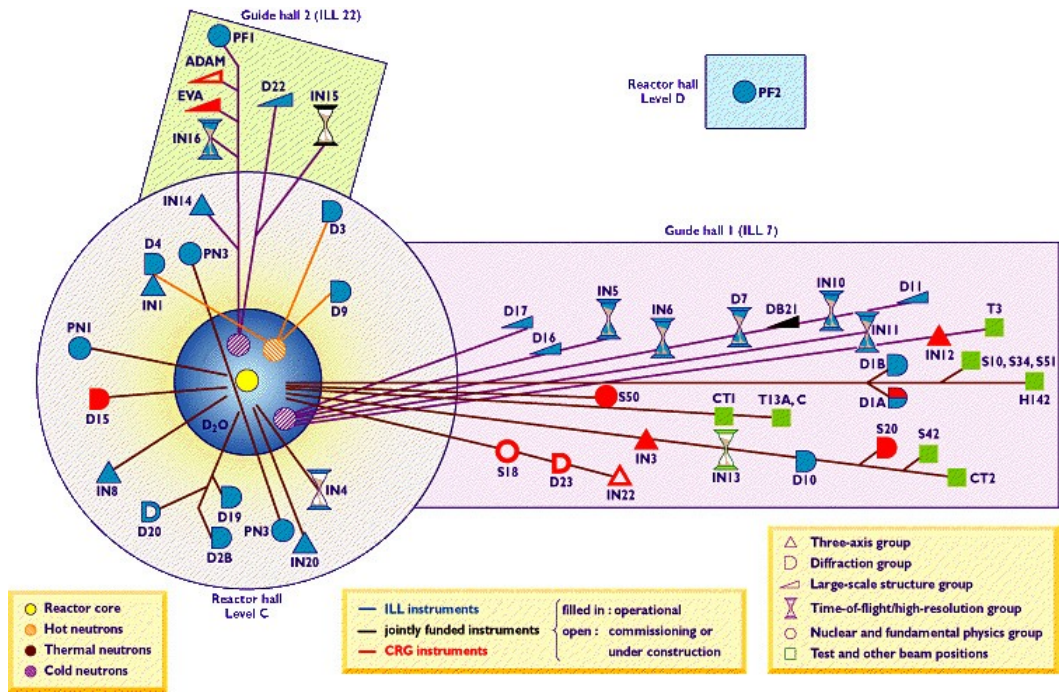

Figure 2.The main view of the beamlines, neutron beam types, and instruments in Institute Laue-LangevinHigh-Flux Reactor.

More detailed view of the cross-section of the ILL reactor is presented in Figure 3.

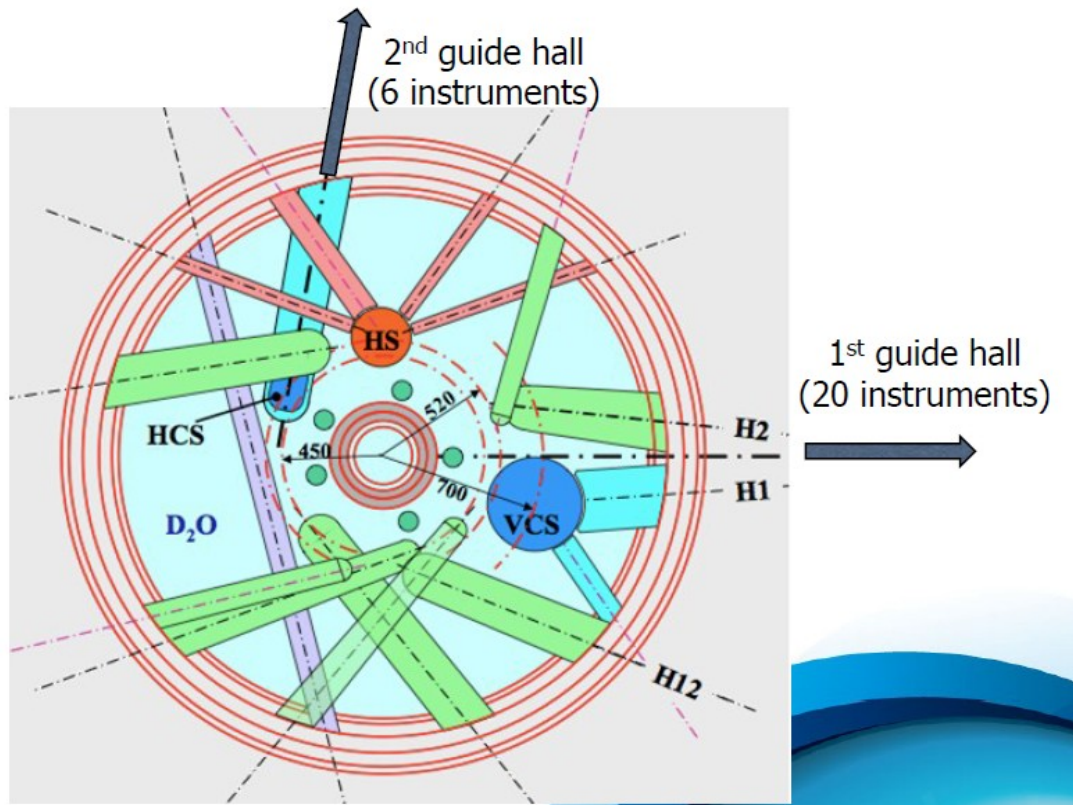

Figure 3.The horizontal cross-section of the ILL reactor. Outer diameter of the tank with D$_2$O is about 2.5 m, HS - hot source, VCS - vertical cold source, HCS - horizontal cold source.

It is interesting to present the dependence of neutron beam flux distributions on distance in beamlines (see Figure 4).

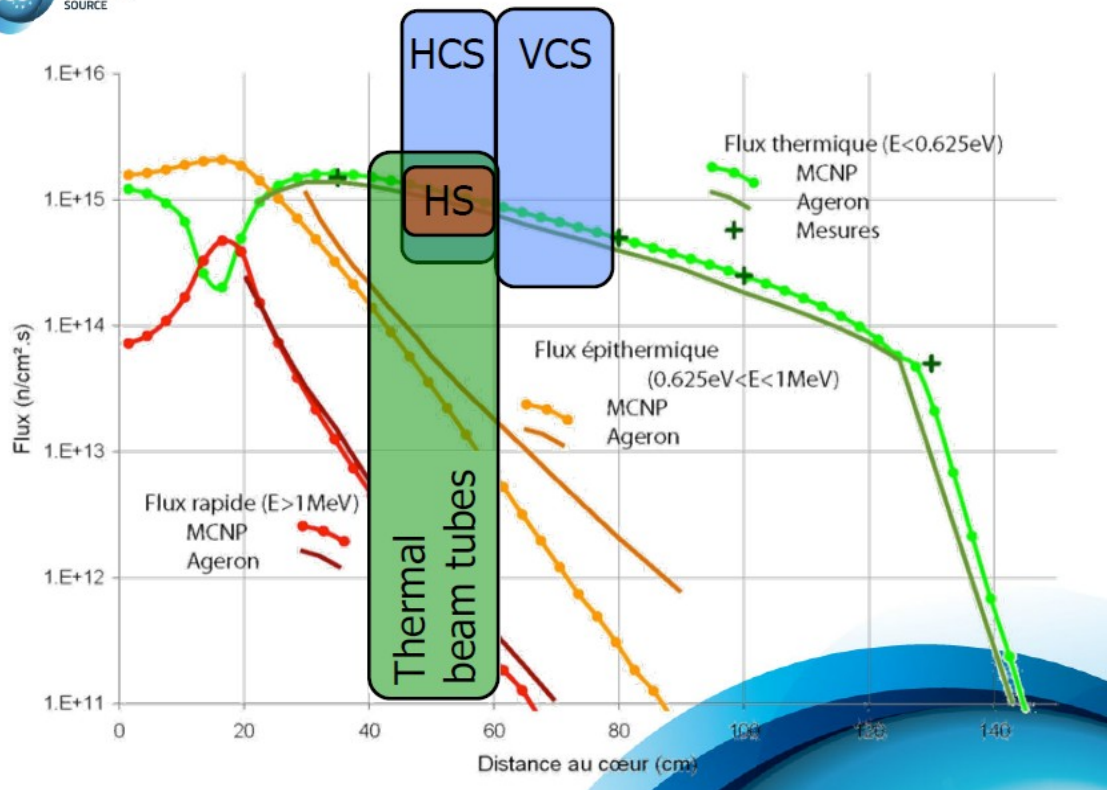

Figure 4. The beam flux distributions dependence on distance in ILL reactor. MCPN - calculation according MCPN code, Ageron - calculations according to [47].

The main parameters of Hi-Flux Advanced Neutron Application ReactorHANARO are following [48]:

Reactor power 30MW  
Coolant                                        Light Water  
Fuel Materials                                 $U_3Si$, 19.75%  
Reflector                                      Heavy water  
Absorber                                       Hafnium  
Max Thermal Flux                               $5 \times 10^{14} n/cm^2/s$  
Typical flux at port nose                      $2 \times 10^{14} n/cm^2/s$  
7 horizontal ports & 36 vertical holes  
The isometric view of the HANARO reactor is presented in Figure 5 [48].

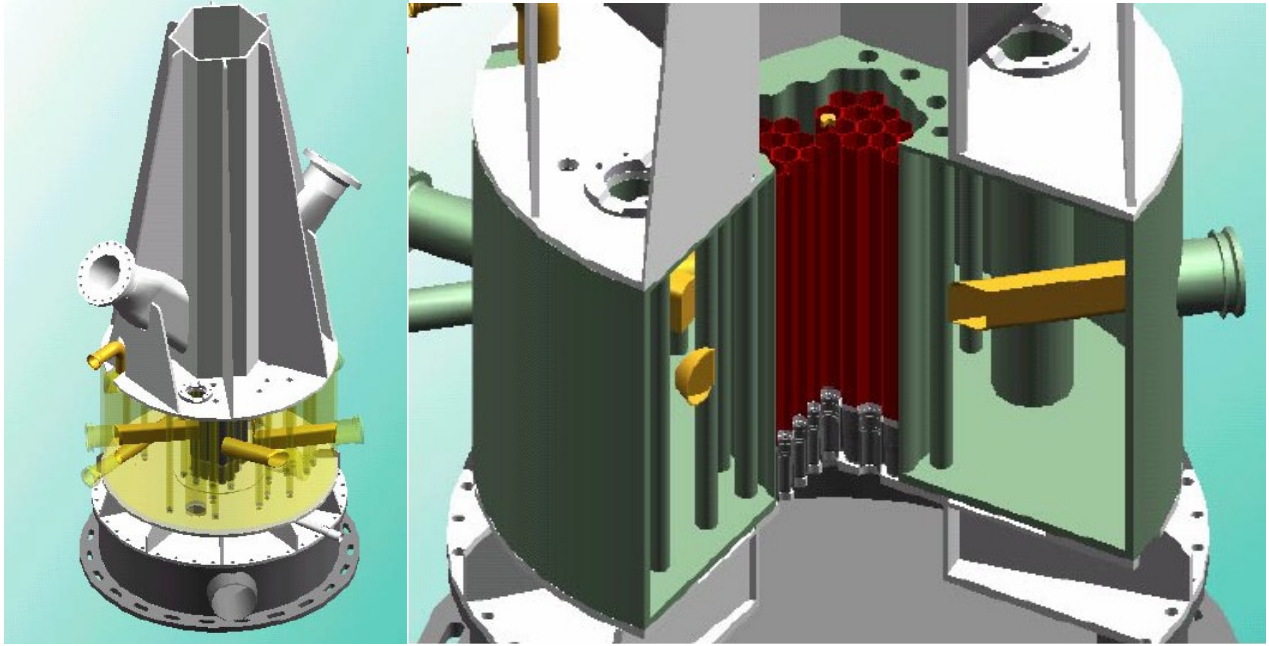

Figure 5. Isometric view of HANARO reactor.

List of tubes and holes, and positions of instruments available at HANARO reactor is presented in Figure 6.

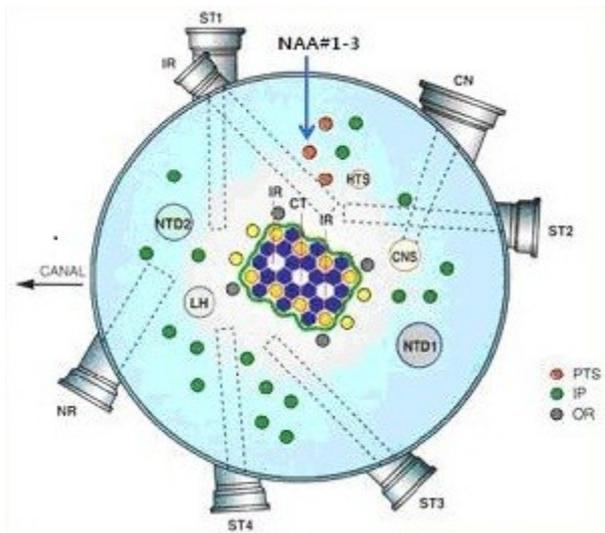

Figure 6. Horizontal section of HANARO reactor with description of instruments positions, tubes and holes. Vertical Holes: IR1 - Fuel test loop, CT, IR2 - Capsule Irradiation and radioisotope (RI) production, OR - Capsule Irradiation and radioisotope production, IP - RI production, HTS -Hydraulic Transfer System for RI Production, PTS - Pneumatic Transfer System for neutron activation, NTD -Neutron Transmutation Doping of Silicon, CNS - Cold Neutron Source. Horizontal Tubes: NR - Neutron Radiography Facility, IR - Ex-core Neutron-irradiation Facility for BNCT and Dynamic Neutron Radiography, ST1 Prompt Gamma Activation Analysis, ST2 High Resolution Power Diffractometer, Four Circle Diffractometer, ST3 - Bio Diffractometer, High Intensity Powder Diffractometer, ST4 - Triple Axis Spectrometer, CN - Cold Neutron Guide.

Calculated with MCNP code thermal neutron flux distribution in horizontal plane is shown in Figure 7 [49]. Maximum unperturbed thermal neutron flux levels (<0.625eV) at the core and reflector regions are estimated to be about $5.03 \times 10^{14}$ and $4.36 \times 10^{14}$ n/cm$^2$/sec, respectively.

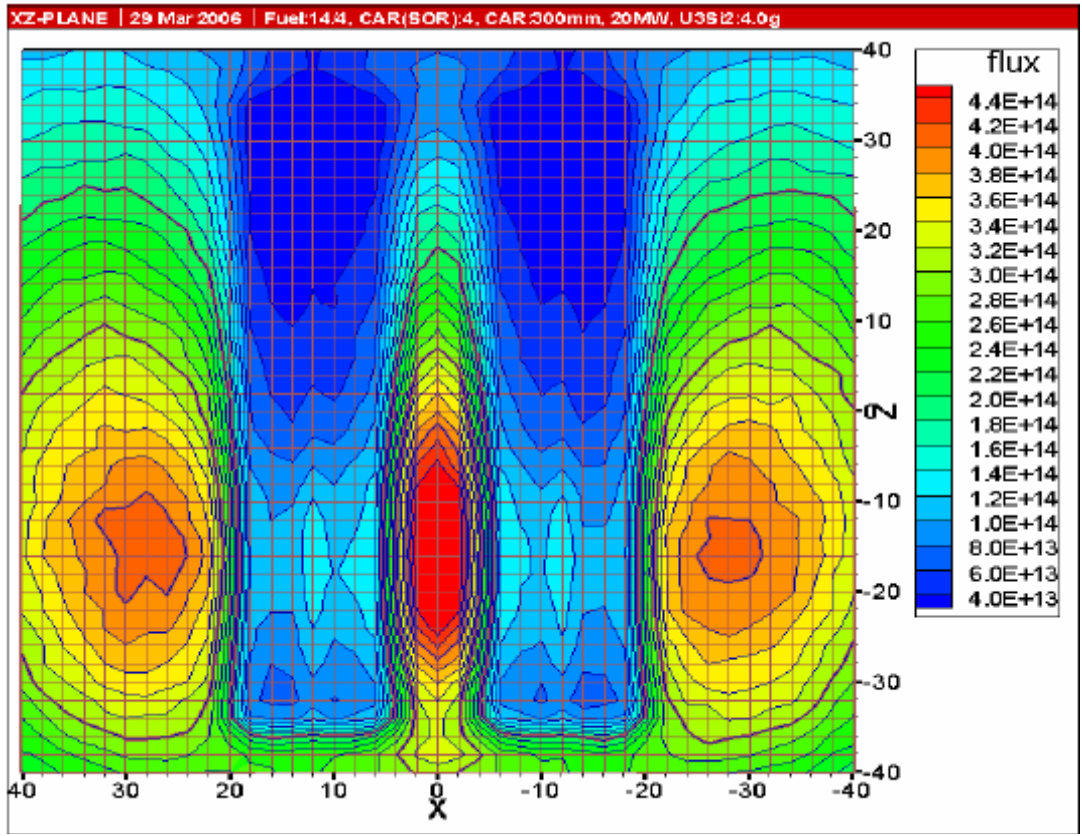

Figure 7. Calculated with MCNP code thermal neutron flux distribution in horizontal plane. Distances in cm, neutron flux is measured in n/cm2/s.

The design characteristics of neutron fluxes at HANARO reactor are presented in Table 2 and Table 3 [50].

Table 2. The perturbed flux at the irradiation holes and beam tubes, at vertical irradiation holes

| Hole ID | Fast neutron flux (>0.821 MeV), $\times 10^{12}$ n/cm$^2$/s | | Thermal neutron flux (<0.625 eV), $\times 10^{14}$ n/cm$^2$/s | |
|---|---|---|---|---|
| | Max | Average | Max | Average |
| CT | 209.752 | 152.699 | 4.393 | 3.125 |
| IR1 | 187.812 | 136.4909 | 3.817 | 2.731 |
| IR2 | 195.172 | 141.481 | 3.928 | 2.811 |
| OR3 | 21.268 | 16.063 | 3.060 | 2.499 |
| IP15 | 2.261 | 1.728 | 1.987 | 1.602 |
| NAA3 | 1.338 | 0.551 | 1.617 | 1.296 |
| HTS | 0.098 | 0.066 | 0.817 | 0.688 |
| NTD1 | 0.070 | 0.055 | 0.393 | 0.328 |
| NTD2 | 0.111 | 0.096 | 0.620 | 0.527 |
| LH | 0.662 | 0.473 | 0.977 | 0.785 |
| CNS | 1.320 | 0.964 | 1.729 | 1.440 |

Table 3. Flux on the ends of beam tubes

| Beam tube | Thermal neutron flux, $\times 10^{14}$ n/cm$^2$/s | Total neutron flux, $\times 10^{14}$ n/cm$^2$/s |
|---|---|---|
| ST1 | 1.892 | 2.262 |
| ST2 | 2.216 | 2.507 |
| ST3 | 2.242 | 2.523 |
| ST4 | 1.532 | 1.691 |
| NR | 0.432 | 0.440 |
| IR | 2.702 | 3.449 |
| CN | 0.968 | 1.004 |

## 1.3. Vibrating wire neutron monitor (VWNM) with composite wires - direct thermal measurements

### 1.3.1. Wires with Gd layer

The $^{157}$Gd has the highest thermal neutron capture cross section of all the stable isotopes in the periodic table. The $^{155}$Gd and $^{157}$Gd isotopes absorb neutrons in a wide range of energies: $^{155}$Gd has 104 resonant levels in range 0.0268-168 eV, and $^{157}$Gd has 60 resonant levels in range 0.0314-307 eV [51]. Thus the $^{157}$Gd capture reaction initiates complex inner-shell transitions that generate prompt $\gamma$-emission displacing an inner core electron. This electron in turn induces internal-conversion (IC) electron emission and finally Auger–Coster–Kronig electron emission along with soft X-ray and photon emissions.

The complete information of neutron capture cross-sections for isotopes of different elements is presented in [52] (see also [53]). For comparison, we note that $^{10}$B has 3840 barn, $^{16}$O has 0.00019 barn, $^{12}$C has 0.0035 barn, $^{1}$H has 0.333 barn, and $^{14}$N has 1.83 barn for neutron capture cross-sections.

The property of gadolinium to capture thermal neutrons is very effectively used in nuclear reactor control rods [54] and shielding of nuclear reactors [55]. Gadolinium is also widely used for industrial neutron radiography with direct exposure technique with gadolinium conversion screen [56]. A typical arrangement is a single 24.4 µm thick Gadolinium metal foil. Gadolinium is used as a core component in the neutron detectors [57, 58]. Gd here serves as a convertor of neutrons to charged particles or photons that finally are measured by gas proportional counters, ionization chambers, scintillation detectors or semiconductor detectors. Exceptionally high absorption of neutrons by Gadolinium was proposed for development of light weight shielding screens by deployment of Gadolinium into the plastic. Such kind of screens containing Gadolinium, boron and hafnium can be helpful for cosmic stations [59] and even in future flights outwards of the Earth as excellent neutron absorbers [60]. $^{157}$Gd is a potential agent for neutron capture cancer therapy [61-63].

### 1.3.2. Gd neutron capture reaction, energy deposition

The Gd isotopes possess very different cross-sections of neutron capture $\sigma$. In Table 4 the cross-section values for thermal neutrons of energy 0.025 eV are presented. [64].

Table 4.

|  | % in natural Gd | cross-section, barn | cross-section, cm$^2$ |
|---|---|---|---|
| natural Gd | 100 | 48890 | 4.889E-20 |
| $^{152}$Gd | 0.2 | 1100 | 1.1E-21 |
| $^{154}$Gd | 2.2 | 90 | 9E-23 |
| $^{155}$Gd | 14.7 | 61000 | 6.1E-20 |
| $^{156}$Gd | 20.6 | 2 | 2E-24 |
| $^{157}$Gd | 15.68 | 255000 | 2.55E-19 |
| $^{158}$Gd | 24.9 | 2.4 | 2.4E-24 |
| $^{160}$Gd | 21.9 | 0.8 | 8E-25 |

The cross-section for $^{157}$Gd is 65 times greater than the thermal neutron capture cross section of $^{10}$B.

The cross-sections of natural Gd and $^{157}$Gd with different energies are presented in Table 5 [65]:

Table 5.

| Wave Length A$^O$ | Energy of Neutrons (eV) | Velocity of Neutrons (m/s) | Capture Cross Section (barns) for natural Gd | Capture Cross Section (barns) for $^{157}$Gd |
|---|---|---|---|---|
| 1 | 0,081894 | 3955 | 13 563.56 | 75 323.47 |
| 1,8 | 0,025276 | 2197,2 | 48 149.41 | 253 778.40 |
| 3 | 0,0090993 | 1318,3 | 70 597.77 | 367 842.60 |
| 4 | 0,0051184 | 988,76 | 89 066.84 | 464 373.40 |

In the first column the neutron wavelength $l = h/mv$ are presented ($h$ - Plank constant, $m$ - 1.67 x 10$^{-24}$ g mass of neutron, $v$ - neutron velocity depicted in the 3rd column).

The two most abundant isotopes of Gd and their corresponding nuclear reactions induced by thermal neutrons are described in the following equations:

$$n + {}^{157}_{64}Gd \rightarrow {}^{158}_{64}Gd^* \rightarrow {}^{158}_{64}Gd + \gamma + conversion \cdot electrons + 7.9 MeV, \sigma_0 = 253929 b \qquad (5)$$

$$n + {}^{155}_{64}Gd \rightarrow {}^{156}_{64}Gd^* \rightarrow {}^{156}_{64}Gd + \gamma + conversion \cdot electrons + 8.5 MeV, \sigma_0 = 60800b \tag{6}$$

Following the absorption of a neutron by the $^{157}$Gd nucleus, for example, several isomeric transitions occur, which result in the release of an average of 3.288 photons. These photons have a wide range of energies with a mean of 2.394MeV. A large number of internal conversion (IC) electrons are emitted owing to the large change in the angular momentum of the low-lying excited states of $^{158}$Gd*. The Gd atom relaxes to the ground state by emitting Auger electrons and characteristic X-rays (see [66]).

Characteristic length of neutrons $L$ in gadolinium is defined by following formula

$$n\sigma L = 1, \tag{7}$$

where $n$ is the Gd atoms density. Using the values of Gd density $\rho$ = 7.9 g/cm$^3$ and atomic mass $A$ = 157.25 g/mol we find $n = \rho/A/N_A$ = 3.03 x 10$^{22}$ cm$^{-3}$.

As it leads from the Table 4 for thermal neutrons (0.025 eV) one can estimate $L$ = 6.7 µm for natural Gd, and $L$ = 1.3 µm for pure $^{157}$Gd. From these estimations, we conclude that by covering vibrating wire with only 10 µm thick layer of natural Gd, all thermal neutrons crossed the wire will be captured.

All components of the neutron capture reaction interact with material around the capturing atom and deposit some energy into it. Exact calculation of the contribution of numerous secondary particles and photons is a complicated task. For electrons and X-ray contributions, we use the value of 70 keV as a first approximation according to [67].

For the estimation of 7.9 MeV gamma-ray contributions we should take into account the wire material properties. Let's consider a Tungsten wire with diameter $d$ = 100 µm, covered with $h$ = 10 µm thickness layer of natural gadolinium. Absorption coefficient $\mu$ for this material is 8.63 x 10$^{-1}$ cm$^{-1}$ ($\mu/\rho$ = 4.47 x 10$^{-2}$ cm$^2$/g [68]). For these values we find that only 68.2 keV of energy will be converted into the heat in 100 µm thick Tungsten.

Both mechanisms of heat contribution lead to $\varepsilon_n$ = 138.2 keV = 2.21 x 10$^{-14}$ J deposition into the vibrating wire made from Tungsten.

### 1.3.3. Composite wire Gd VWNM, frequency/temperature dependence

[69] As a first approximation of Gadolinium VWNM let's consider a simplest one-wire model of vibrating wire monitor with following parameters: $d_W$ - base wire diameter, $l_W$ - wire length, $h$ - Gd layer thickness, $l_A$ - monitor aperture free for neutron flux (indeed some neutrons can penetrate to the wire through magnet system parts, which are neglected here). The vibrating wire total diameter $d = d_W + 2h$, cross section of base wire $S_W = \pi d_W^2/4$, cross-section of Gd layer $S_{Gd} = \pi h(d_W + h)$, volume of base wire $V_W = S_W l_W$, volume of Gd $V_{Gd} = S_{Gd} l_W$.

Also we introduce $\rho_W$ -density of base wire material, $\rho_{Gd}$ - density of Gd layer, so the mean density of compose wire is $\rho = \rho_W \dfrac{1 + \rho_{Gd} V_{Gd}/\rho_W V_W}{1 + V_{Gd}/V_W}$. \hfill (8)

Second harmonic frequency at temperature $T_0$ is found from the formula:

$$F_0 = \frac{1}{l_W}\sqrt{\sigma_0/\rho}, \tag{9}$$

where $\sigma_0$ - wire initial tension. This tension can be presented as

$$\sigma_0 = \frac{f_W^0 + f_{Gd}^0}{S_W + S_{Gd}} \tag{10}$$

where $f_W^0$ is tension force of base wire and $f_{Gd}^0$ is tension force of Gd layer.

Tension procedure can be presented as a stretching of the unstressed wire initial length $l_0$ to the length $l$ with fixing of wire ends afterwards. We suppose that this procedure is done at initial temperature of all parts of VWNM $T_0$. According to Hook's law we have

$$f_W^0 = \frac{l-l_0}{l_0} E_W S_W = \frac{\Delta l_0}{l_0} E_W S_W, \qquad (11)$$

$$f_{Gd}^0 = \frac{l-l_0}{l_0} E_{Gd} S_{Gd} = \frac{\Delta l_0}{l_0} E_{Gd} S_{Gd}, \qquad (12)$$

where $E_W$ and $E_{Gd}$ are wire material and Gd modulus's of elasticity

At the temperature $T$ (change $\Delta T = T - T_0$ in assumption that $T$ is mean temperature of the wire) we have

$$f_W^T = \frac{l-l_0(1+\alpha_W(T-T_0))}{l_0} E_W S_W = \frac{\Delta l_0 - l_0 \alpha_W \Delta T}{l_0} E_W S_W = f_W^0 - \alpha_W E_W S_W \Delta T, \qquad (13)$$

$$f_{Gd}^T = \frac{l-l_0(1+\alpha_{Gd}(T-T_0))}{l_0} E_{Gd} S_{Gd} = \frac{\Delta l_0 - l_0 \alpha_{Gd} \Delta T}{l_0} E_{Gd} S_{Gd} = f_{Gd}^0 - \alpha_{Gd} E_{Gd} S_{Gd} \Delta T, \qquad (14)$$

where $\alpha_W$ - wire material coefficient of thermal expansion and $\alpha_{Gd}$ - the same for Gd layer. As a result we can find the VWM composite wire tension $\sigma^T$ at mean temperature of the wire $T$:

$$\sigma^T = \frac{f_W^T + f_{Gd}^T}{S_W + S_{Gd}} = \frac{f_W^0 + f_{Gd}^0 - (\alpha_W E_W S_W + \alpha_{Gd} E_{Gd} S_{Gd}) \Delta T}{S_W + S_{Gd}} = \sigma_0 - \frac{\alpha_W E_W S_W + \alpha_{Gd} E_{Gd} S_{Gd}}{S_W + S_{Gd}} \Delta T. \qquad (15)$$

We suppose that only the wire temperature is changed and all the other parts of the VWNM remains at the initial temperature $T_0$.

From (9) and (15) it is lead

$$\frac{\Delta F}{F_0} \approx \frac{\Delta \sigma}{2\sigma_0} \approx -\frac{\alpha_W E_W S_W + \alpha_{Gd} E_{Gd} S_{Gd}}{2\sigma_0 (S_W + S_{Gd})} \Delta T \qquad (16)$$

(we neglect all the geometrical changes due to the temperature change except the tension term in (9)).

And thereby

$$\frac{\Delta F}{\Delta T} = -\frac{(\alpha_W E_W S_W + \alpha_{Gd} E_{Gd} S_{Gd}) F_0}{2\sigma_0 (S_W + S_{Gd})} \qquad (17)$$

(compare with uniform wire formula $\frac{\Delta F}{\Delta T} = \frac{E \alpha F_0}{2\sigma_0}$ presented in [70]).

In case of Gd VWNM wire overheating arises as a result of neutron beam capture in the wire. The total power heating in the wire can be represent as

$$W = I_n \varepsilon_n l_A (d+2h), \qquad (18)$$

where $I_n$ - is neutron flux, $\varepsilon_n$ - is deposition of energy of a single captured neutron, $l_A(d+2h)$ - is a cross section of wire capturing aperture. Let's suppose that as a result of heating a triangle profile of temperature in the wire is set. More accurate estimations [71] show that the triangle model of temperature profile is a good approximation.

For triangle profile of balanced temperature of the wire (max temperature $T_M$ about ambient temperature $T_0$) we have sink of power via thermal conduction process in composite wire:

$$W_\lambda = \frac{8\Delta T}{l_W}(\lambda_W S_W + \lambda_{Gd} S_{Gd}), \tag{19}$$

here $\Delta T = (T_M - T_0)/2 = T - T_0$ is mean value of wire overheating, $\lambda_W$ and $\lambda_{Gd}$ are thermoconductivity coefficients of base wire and Gd correspondingly.

The thermal sink produced by thermal radiation process is

$$W_{RAD} = \varepsilon \sigma_{ST\_B} T^4 \pi(d+2h)l_W - \varepsilon \sigma_{ST\_B} T_0^4 \pi(d+2h)l_W \approx 4\varepsilon \sigma_{ST\_B} T_0^3 \Delta T \pi(d+2h)l_W, \tag{20}$$

with the assumption that wire emissivity $\varepsilon$ is the same for thermal emission and deposition.

In case wire is placed in the air, there is a convection thermal sink also:

$$W_{CONV} = \Delta T \alpha_{CONV} \pi(d+2h)l_W, \tag{21}$$

where $\alpha_{CONV}$ is the coefficient of convective losses.

Balance temperature of the wire $T$ is found by noting the fact that the power $Q$ deposited into the wire is equal to the sum of all thermal sinks:

$$W = W_\lambda + W_{RAD} + \eta W_{CONV}, \tag{22}$$

where $\eta = 1$ if VWM is placed in air and $\eta = 0$ if VWM is placed in vacuum.

From this equation we find

$$W/\Delta T = 8(\lambda_W S_W + \lambda_{Gd} S_{Gd})/l_W + 4\varepsilon \sigma_{ST\_B} T_0^3 \pi d l_W + \eta \alpha_{CONV} \pi d l_W / 2. \tag{23}$$

Equation (23) coupled with Eqs. (17) and (19) defines the relation between vibrating wire frequency shift and value of measured neutron flux:

$$\frac{\Delta F}{I_n} = -\frac{(\alpha_W E_W S_W + \alpha_{Gd} E_{Gd} S_{Gd}) F_0 \varepsilon_n l_A (d+2h)}{2\sigma_0 (S_W + S_{Gd})[8(\lambda_W S_W + \lambda_{Gd} S_{Gd})/l_W + 4\varepsilon \sigma_{ST\_B} T_0^3 \pi d l_W + \eta \alpha_{CONV} \pi d l_W / 2]}. \tag{24}$$

Our long time experience on vibrating wire resonators operation shows that the measurement accuracy of physical process which results in frequency shift, can be estimated to be on the order of 0.01 Hz. For convection coefficient we introduce experimentally measured value of 380 W/m$^2$/K [72].

**1.3.4. Gd VWNM response time**

To the first approximation the response time of the VWNM can be found from the following thermal equation

$$W\tau = l_W(S_W \rho_W c_W + S_{Gd} \rho_{Gd} c_{Gd})\Delta T, \tag{25}$$

where $c_W$ is the specific heat coefficient of the base wire and $c_{Gd}$ is specific heat coefficient of the Gd. Equation (25) is a simple relation that allows us to estimate how much time is needed to increase the temperature of the wire at the balanced value $\Delta T$. Taking into account the equation (23) we find

$$\tau = \frac{l_W(S_W \rho_W c_W + S_{Gd} \rho_{Gd} c_{Gd})}{8(\lambda_W S_W + \lambda_{Gd} S_{Gd})/l_W + 4\varepsilon \sigma_{ST\_B} T_0^3 \pi d l_W + \eta \alpha_{CONV} \pi d l_W / 2}. \tag{26}$$

For VWM with parameters defined in the previous section (base wire diameter from Tungsten - 100 µm, natural Gd layer thickness is 10 µm, wire total length 40 mm) we find $\tau$ =0.3 s for VWNM working in air and $\tau$ =3 s in case of vacuum.

### 1.3.5. Middle sizes scale VWNM

To evaluate the corresponding accuracy of temperature (according to Eq. (17)) and the neutron flux, we consider the VWNM with following parameters: base wire diameter made from Tungsten is 100 µm, natural Gd layer thickness is 10 µm, wire total length 40 mm and aperture 20 mm. Monitors with approximately these scale of parameters we call middle size scale VWNM. The main view of such type monitor is presented in Figure 8.

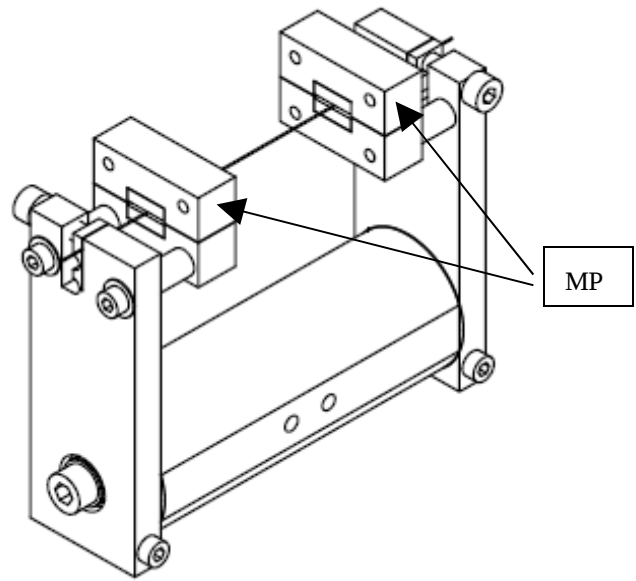

Figure 8. Main view of VWNM with aperture 40 mm (the space between magnet poles (MP)).

Frequency resolution 0.01 Hz corresponds to accuracy in temperature about 1.8 mK. In addition, we find that the accuracy in the neutron flux measurement is about $2 \times 10^{10}$ n/cm$^2$/s if vibrating wire is set in air, and $10^9$ n/cm$^2$/s in case of vacuum, respectively. The temperature increase of wire can reach few hundreds K (see [73]) so dynamic ranges of detectable neutron flux intensity are ca $2 \times 10^{10}$ - $2 \times 10^{15}$ n/cm$^2$/s for air (response time according to (26) ca 0.3 s) and ca $10^9$ - $10^{14}$ n/cm$^2$/s for vacuum (response time ca 3 s). Such type of VWNM can be used for profiling of neutron beams with beam sizes about few tens of mm.

### 1.3.5. Small sizes scale VWNM

As one can see from Eqs. (24) and (26), the resolution and response time of the VWNM depends essentially on the wire length and diameter. As reducing these parameters, the level of resolution and response time can also be reduced. For tungsten wire with ca 10 µm diameter and natural Gd layer with ca 2 µm thickness, a wire with total length ca 5 mm can be completely covered by a permanent magnet and the aperture can be defined also as ca 5 mm. Monitors with approximately these scale of parameters we call small size scale VWNM.

Corresponding to 0.01 Hz frequency resolution, the temperature resolution is 0.03 mK and the dynamic ranges of detectable neutron flux intensity are ca $3 \times 10^8$ - $3 \times 10^{13}$ n/cm$^2$/s for air (response time ca 1ms) and ca $8 \times 10^7$ - $8 \times 10^{12}$ n/cm$^2$/s for vacuum (response time ca 70 ms), respectively. Such type of VWNM may be used in field of nuclear safety as a portable, high-speed environmental monitor.

### 1.3.6. Gd deposition on the base wire

Metallic Gadolinium is available on the market (see e.g. [74]). As a pure metal it is stable in a dry atmosphere but forms an oxide coating when exposed to moist air. It reacts slowly with water and is soluble in acids. So it seems that the better way will be use of more mechanically stable materials like tungsten as a base wire (modulus of elasticity of W is 411 GPa, compared with 56.2 GPa for Gd, which is important for good mechanical resonator characteristic, and tensile strength is equal to 1920 MPa for W and just only 193 MPa for Gd). In addition, the electrical resistivity of W is much lower (5.4 µOhm cm for W and 134 µOhm cm for Gd), which makes easier to process oscillations by using special electronics unit.

At our disposal we have gadolinium oxide powder and have a technology to make soluble compounds of gadolinium: $GdCl_3$ and $GdBr_3$. Our intention is to provide electrodeposition of Gd on tungsten wire by using electrolytes base on these compounds.

### 1.3.6. Gd degradation

Each neutron capture by gadolinium isotopes with large cross-sections ($^{155}Gd$ and $^{157}Gd$) convert them to the practically "dead" ones ($^{156}Gd$ and $^{158}Gd$ correspondingly). The number of all Gd atoms in the wire aperture volume is $2 \times 10^{18}$ for typical middle sizes scale VWNM. So the life time at the intensity level of 1000 times of the solvable flux is about 60 days. For small sizes scale VWNM, the life time of the monitor is estimated more than one year due to lower resolution flux.

### 1.4. "Resonant Target Vibrating wire neutron monitor (RT-VWNM)

For the measurements of neutron beam gradients, method suggested in [9] can be used. The core of the method consists in the measurements of particles/radiation arising from neutron beam scattering at vibrating wire in synchronism with opposite positions of wire in mechanical oscillation process. In Figure 9 the main view of the concept is presented. The first harmonic oscillations are generated in the plane orthogonal to the beam presented by ellipse with graded color corresponding to beam density. Wire at two extreme positions during the oscillations is drawn.

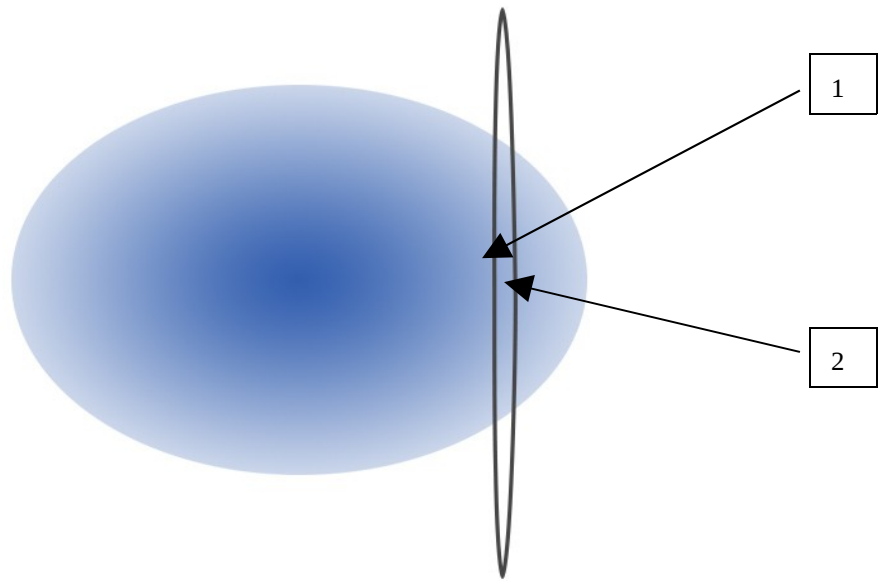

Figure 9.Vibrating wire in density gradient area of beam. In position (1) more particles penetrate the wire than in position (2). Difference of penetrating particles forms signal proportional to beam density gradient.

In case of precise mechanical oscillations the differential signal of such measurements provides direct signal of the spatial gradient of the neutron flux. The method can be useful in the presence of high level background in the range of measured particles/radiation. Measurements need times less than one ms (oscillation frequency is normally few kHz). Compared with few sec measurements in thermal deviation measurement described above, this can be an advantage of the proposed method. Measurements should be complemented by acquisition of absolute signals coming from scattering and additional sampling of wire oscillation frequency representing the temperature deviation of the wire.

In [9] the method was tested with laser beam where the scattered photons were measured by fast photodiodes at opposite maximal deviations of the wire from straight-line position. A stainless steel wire with 100 μm diameter and 80 mm length was used, and it oscillates with stabilized amplitude about 200 about μm (peak to peak). Results of the experiment are presented in Figure 10. In this Figure we see three different approaches to the beam profiling. Blue and light blue lines presented the measurements of vibrating wire frequency depend on heat disposal from the measured beam and environmental conditions lead to signal drifts. The absolute signal (red and magenta lines) does not require the oscillation process and can be obtained even at no vibrating wire. However the signal of measured beam is added to some background. In our case the value of background is about 500 mV and excess of useful signal is quite high – about

2000 mV, but in other cases background can be significantly larger. In resonant target method the background is completely removed by subtracting procedure and outside of the beam is equal to zero (green and brown lines).

In the described experiment we used home-made electronics with measurement gates and delays, correlated with frequency of the wire oscillations. This precise measurement technique allowed obtaining accuracy of measurements in the range of mHz within 1 s of sampling interval. Another possibility is use of special electronic devices to extract signal in narrow gap around the oscillation frequency – so called lock-in amplifier with unprecedentedly small filtering gap [75].

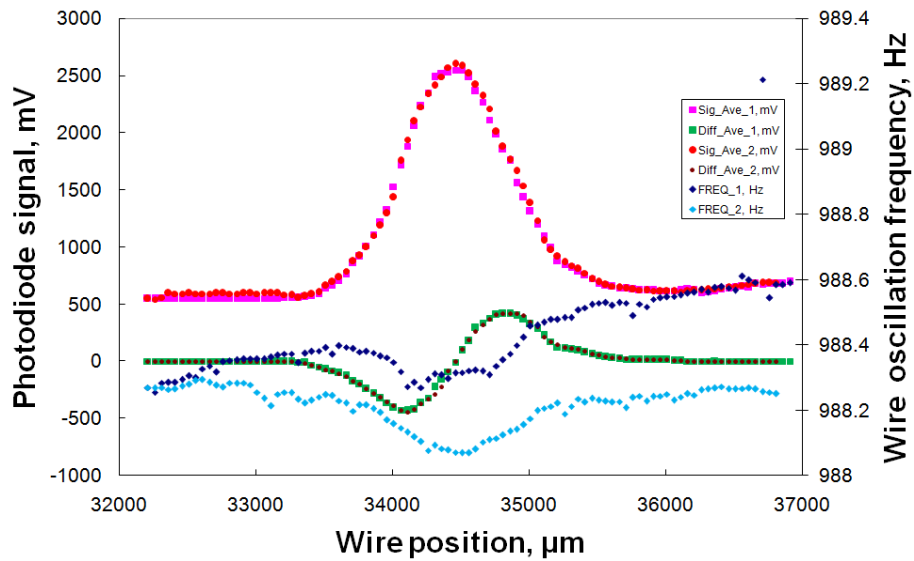

Figure 10. Scan of laser beam as function of oscillating wire position. Red and magenta lines - averaged absolute signal from photodiode (forward and backward scan), green and brown lines - differential signal from photodiode (forward and backward scan), dark blue and light blue lines - frequency signal (forward and backward).

### 1.4.1. Wire

The core piece of the proposed method is the wire. Besides high quality mechanical oscillations with very stable amplitude it should also provide enough quantity of secondary particles/radiation for measurements. It seems that more adequate choice is use of the same composite wires with Gd layer. Prompt gamma-rays arising during the scattering can be measured by common techniques of gamma-rays registration.

Another very important characteristic of the wire itself is its length. For real spatial measurements of the neutron beam the aperture of VWNM should be at least of order of the beam size of neutron guide tube. Normally it is few tens of cm so we have to use at least twice longer wires taking into account that some parts of the oscillating wire is covered by permanent magnets needed for oscillation generation. Maximal length that we used before was 80 mm so we need to develop vibrating wire resonators with longer lengths.

### 1.4.2. Gamma-ray detector for RT-VWNM

At the first stage of development of RT-VWNM we intend to use widely used sodium iodide (NaI) scintillator based detectors. Such detectors have good speed of response (ns range) so the serial measurement of few hundreds of μs with few kHz duty factor is a feasible task.

A simpler possibility is use of ultraviolet PIN photodiodes, which also react on gamma-rays with some sensitivity (see e.g. [76, 77]).

Sensitivity of RT-VWNM is actually defined by characteristics of these gamma-ray detectors but it seems that they provide possibility to detect even neutron beams of radioisotope neutron sources such as $^{241}$Am/Be or $^{252}$Cf.

### 1.5. Conclusion

The main advantage of the proposed instruments VWNM and RT-VWNM is that they are very robust, and that the radiation resistive mechanical resonator consists only from wire ends clips, magnet poles with permanent magnets, and mechanically stable wires. Therefore, the vibrating wire-based sensors have capability to operate in hard conditions (high operational, power and temperature cycling, thermal shock, thermal storage, autoclave, electromagnetic irradiation fields

[78]). The important advantages of properly constructed vibrating wire monitors are inherent long-term stability, negligible zero-drift, high precision and resolution, good reproducibility and small hysteresis. Another advantage of the vibrating wire sensors is that the frequency signal is imperturbable and can be transmitted over long cable with no loss or degradation of the signal. Taking into account that RT-VWNM and VWNM modifications cover a wide range of measured neutron beam flux intensities we can confidently state that the proposed vibrating wire based neutron monitors can be widely used for all applications of neutron beams. VWNM as a precise monitor with excellent spatial resolution for high flux neutron beams of the specialized neutron sources with multibranch infrastructure of numerous instruments for material research. Small sizes scale VWNM may be used in the field of nuclear safety as a portable, high-speed environmental monitor. RT-VWNM as a robust and reliable instrument with also excellent spatial resolution can be applied for low flux neutron beam diagnostics (e.g. for centers of neutron therapy). Specialized multiwire VWNM with capability of rotating along the beam axis can be used for the recovery of complicated 2D profiles of large cross-section neutron beams in neutron tomography, imaging and radiography. VWNM's can be used in 18 MeV Cyclotron (Cyclone 18) of Yerevan's oncological center for direct beam profile measurements in medical treatment. Another area of usage can be diagnostics of neutron beam planned to be generated at Cyclone-18 [16]. Preliminary experiments and tests are planned to perform on the $^{252}$Cf spontaneous fission neutron source accessible in Yerevan Physics Institute.